\documentclass[aip,jcp,11pt]{revtex4-1}

\draft
\usepackage{amsmath}
\usepackage{amssymb}
\usepackage{graphicx}
\usepackage{esint}
\PassOptionsToPackage{version=3}{mhchem}
\usepackage{mhchem}
\usepackage[unicode=true,pdfusetitle,
 bookmarks=true,bookmarksnumbered=false,bookmarksopen=false,
 breaklinks=false,pdfborder={0 0 1},backref=section,colorlinks=false]
 {hyperref}

\makeatletter

\usepackage{amsthm}

\usepackage{tikz}\usetikzlibrary{arrows}
\usepackage{caption}
\usepackage{subcaption}
\usepackage{color}

\usepackage{multirow}\usepackage{bigdelim}

\usepackage[section]{placeins}

\newcommand{\bra}[1]{\langle #1 \mid}
\newcommand{\ket}[1]{\mid #1 \rangle}
\newcommand{\bracket}[2]{\langle #1 \mid #2 \rangle}
\usepackage{bbold}\newcommand{\identity}{\mathbb{1}}

\newcommand{\Q}{\mathbf{Q}}
\renewcommand{\P}{\mathbf{P}}
\DeclareMathOperator\erf{erf}

\renewcommand{\H}{\mathbf{H}}
\renewcommand{\S}{\mathbf{S}}
\renewcommand{\U}{\mathbf{U}}
\newcommand{\E}{\mathbf{E}}
\renewcommand{\C}{\mathbf{C}}

\newcommand{\matr}[1]{\mathbf{#1}}

\usepackage{mathtools}

\renewcommand{\imath}{\text{\rm{i}}}

\usepackage{titlesec}\titleformat{\chapter}
{\normalfont\huge\bfseries}{\chaptertitlename\ \thechapter:\ \ }{0em}{} 
\titlespacing*{\chapter}{0pt}{0pt}{30pt}

\begin{document}

\author{Alexander Humeniuk}
\author{Roland Mitri\'{c}}
\affiliation{Institut f\"{u}r Physikalische und Theoretische Chemie, Julius-Maximilians Universit\"{a}t W\"{u}rzburg, Emil-Fischer-Stra\ss e 42, 97074 W\"{u}rzburg}
\email{roland.mitric@uni-wuerzburg.de}
\title{Fully quantum non-adiabatic dynamics in electronic-nuclear coherent state basis}



\makeatother


\begin{abstract}
Direct dynamics methods using Gaussian wavepackets have to rely only 
on local properties, such as gradients and hessians at the center 
of the wavepacket,
so as to be compatible with the usual quantum chemistry methods.
Matrix elements of the potential energy surfaces between wavepackets therefore usually have to be approximated\cite{bat_approximation}. 

It is shown, that if a modified form of valence bond theory is used instead of the usual MO-based theories, the matrix elements can be obtained exactly. This is so because the molecular Hamiltonian only contains the Coulomb potential, for which matrix elements between different basis functions (consisting of Gaussian nuclear and electronic orbitals) are all well-known. In valence bond theory the self-consistent field calculation can be avoided so that the matrix elements are analytical functions of the nuclear coordinates. 

A method for simulating non-adiabatic quantum dynamics is sketched, 
where coherent state trajectories are propagated ``on the fly'' on adiabatic potential energy surfaces without making approximations to the matrix elements responsible for the coupling between trajectories.
\end{abstract}
\maketitle

\section{Introduction}
The CCS method\cite{CCS_method, xray_ccs} needs global knowledge of potential energy surfaces, while most quantum chemistry methods only provide local information such as adiabatic energies, gradients and non-adiabatic couplings for a specific nuclear geometry.
Many algorithms for simulating (semi-classical) molecular dynamics (MD) are designed to rely only on this limited information, which is requested ``on the fly'' as the nuclear trajectory is propagated in time.
For the purpose of the program that drives the MD simulation, the electronic structure methods are considered black boxes that give the local properties of the energy surface, so that the same MD program can be combined with different methods.

The determinantion of the electronic structure usually starts with a self-consistent field (SCF) calculation which produces orthogonal orbitals.  By occupying the lowest of these orbitals one or more reference states are defined. On top of this electron correlation and electronic excitations are added by perturbation theory, linear response calculations or some sort of configuration interaction. These approaches based on orthogonal but delocalized molecular orbitals are quite efficient and give access to complex potential energy surfaces. 

Nevertheless, one should not forget that the underlying molecular hamiltonian is quite simple, it only contains the Coulomb interaction between charged particles. Valence bond methods, which avoid the self-consistent field calculation altogether, open a route to global potential energy surfaces. Matrix elements of valence bond structures, which are built by pairing electrons in atomic orbitals, are analytical functions of the nuclear coordinates. Valence bond structures can be expanded into non-orthogonal Slater determinants and if Gaussian orbitals are used to represent both the single-particle orbitals for electrons and nuclei (in the spirit of nuclear electronic orbital method\cite{neo,nbo}), the matrix elements of those can be expressed in closed form. 
The Born-Oppenheimer separation is thus avoided, since electrons and nuclei are treated on a similar footing.

The wavefunction is expanded into a moving basis set. Each moving basis vector consists of a nuclear coherent state, that moves on an adiabatic potential energy surfaces and can hop stochastically between different electronic states. 
These adiabatic electronic eigenstates are obtained by configuration interaction between valence bond structures (or non-orthogonal Slater determinants). 

Since SCF calculations are avoided, one can also calculate matrix elements of the hamiltonian between moving basis vectors differing in their nuclear coordinates, which is not possible for quantum chemistry methods based on MO theory. From the coupling between different basis vectors, equations of motion for the expansion coefficients can be derived. This allows to include interference effects between trajectories with different phases.

The time-evolution of the wavefunction becomes exact if the electronic basis set approaches completeness and if the swarm of trajectories is large enough to cover the relevant parts of the nuclear phase space. 

The time needed to evaluate the hamiltonian between moving basis vectors is dominated by the time necessary for matrix elements of the electronic valence bond structures, $t_{\text{VB}}$. The time needed to propagate $N_{\text{traj}}$ uncoupled trajectories will therefore scale as $N_{\text{traj}} \times t_{\text{VB}}$, while the time for propagating the coupled equations will scale as $N_{\text{traj}}^2 \times t_{\text{VB}}$.

Valence bond theory is usually not the method of choice for molecular dynamics simulations due to its prohibitive scaling with the system size. However, the incorporation of nuclear quantum effects by means of interacting trajectories turns out to be much easier than with MO based methods. The method presented here will also benefit from future improvements in valence bond theory. 

\section{Method Description}
\subsection{Matrix Elements}
The starting point is the full molecular Hamiltonian for $N_{\text{at}}$ atoms with $N_{\text{elec}}$ electrons in an external time-dependent electric field $\mathbf{E}(t)$ (a real-valued 3-dimensional vector):

\begin{equation}
\begin{split}
\hat{H} = &- \sum_{a=1}^{N_{\text{at}}} \frac{1}{2 M_a} \nabla_a^2 + \sum_{a=1}^{N_{\text{at}}} \sum_{b > a}^{N_{\text{at}}} \frac{Z_a Z_b}{\vert \mathbf{R}_a - \mathbf{R}_b \vert} + \mathbf{E}(t) \cdot \sum_{a=1}^{N_{\text{at}}} Z_a \mathbf{R}_a \\
 &- \sum_{k=1}^{N_{\text{elec}}} \frac{1}{2} \nabla_k^2 + \sum_{k=1}^{N_{\text{elec}}} \sum_{k > l}^{N_{\text{elec}}} \frac{1}{\vert \mathbf{r}_k - \mathbf{r}_l \vert} - \mathbf{E}(t) \cdot \sum_{k=1}^{N_{\text{elec}}} \mathbf{r}_k \\
 & -\sum_{k=1}^{N_{\text{elec}}} \sum_{a = 1}^{N_{\text{at}}} \frac{Z_a}{\vert \mathbf{r}_k - \mathbf{R}_a \vert}
\end{split}
\end{equation}
$M_a$ denotes the mass of the nucleus $a$, $Z_a$ is its nuclear charge and $\mathbf{R}_a$ and $\mathbf{r}_k$ are the coordinate vectors of the nucleus $a$ and the electron $k$, all in atomic units. 
The dipole approximation is made for treating the interaction with the external field $\mathbf{E}(t)$. 
The hamiltonian can be separated into three parts:
\begin{equation}
\hat{H} = \hat{H}_{\text{nuc}}\left(\left\{ \nabla_a, \mathbf{R}_a \right\}_{a=1,\ldots,N_{\text{at}}}\right) + \hat{H}_{\text{elec}}\left(\left\{ \nabla_k, \mathbf{r}_k \right\}_{k=1,\ldots,N_{\text{elec}}} \right) + \hat{H}_{\text{elec-nuc}}\left(\left\{ \mathbf{R}_a \right\}_{a=1,\ldots,N_{\text{at}}}, \left\{ \mathbf{r}_k \right\}_{k=1,\ldots,N_{\text{elec}}} \right) 
\end{equation}
The first part only depends on the nuclei, the second only on the electrons and the third accounts for the attractive interaction between nuclei and electrons.

Now a basis will be introduced both for the electronic and the nuclear degrees of freedom. 

\paragraph{\textbf{Nuclear basis.}}
Atoms are treated as distinguishable particles although atoms of the same type are strictly speaking indistinguishable and obey either Fermi or Bose statistics. With this approximation a nuclear coherent state is the product of the coherent states for each individual atoms:

{\small
\begin{equation}
\bracket{\mathbf{R}_{a},\ldots,\mathbf{R}_{N_{\text{at}}}}{\Q_1,\P_1;\ldots;\Q_{N_{\text{at}}},\P_{N_{\text{at}}}} =
      \prod_{a=1}^{N_{\text{at}}} 
          \left( \frac{2 \alpha_a}{\pi} \right)^{3/4} \exp\left(
              -\alpha_a \left(\mathbf{R}_a - \Q_a \right)^2 
              + \imath \P_a \cdot \left(\mathbf{R}_a - \Q_a \right) 
              + \frac{\imath}{2} \P_a \cdot \Q_a \right)
\end{equation}
}

The width parameter $\alpha_a$ of the nuclear wavepacket for each atom is taken to be proportional to the nuclear mass:
\begin{equation}
\alpha_a = \alpha_{H} \frac{m_a}{m_H}
\end{equation}
For the width parameter of hydrogen, we set $\alpha_H = 12.5$.

An atomic nuclear coherent state can be written as an s-type Gaussian orbital $g_{1s}$, whose center is a complex vector $\mathbf{A}_a$:
{\small
\begin{equation}
\bracket{\mathbf{R}_{a}}{\Q_a,\P_a}  = \exp\left( \frac{\imath}{2} \P_a \cdot \Q_a - \frac{1}{4 \alpha_a} \P_a^2 \right) \times \underbrace{
      \left( \frac{2 \alpha_a}{\pi} \right)^{3/4} 
      \exp\left( - \alpha_a \left(\mathbf{R}_a - \underbrace{\left[ \Q_a + \frac{\imath}{2 \alpha_a} \P_a \right]}_{\text{complex vector} \mathbf{A}_a} \right)^2 \right)}_{g_{1s,\alpha_a}\left(\mathbf{R}_a - \mathbf{A}_a\right)}
\end{equation}
}
Therefore the matrix elements for nuclear orbitals can be found using very similar formulae \cite{szabo_ostlund} as for electronic orbitals. The matrix elements for the attractive interaction $\hat{H}_{\text{elec-nuc}}$ between electrons and nuclei will reduce to formulae similar to the usual electron repulsion integrals. 

Instead of labeling the coherent states by their position and momenta, $\Q_i$ and $\P_i$, the two can be combined into a complex vector
\begin{equation}
\mathbf{z}_{a,i} = \sqrt{\alpha_a} \Q_{a,i} + \imath \sqrt{\frac{1}{4 \alpha_a}} \P_{a,i}
\end{equation}
and we will use the notations, $\ket{\mathbf{z}_i} = \ket{\Q_i,\P_i}$ interchangeably.

\paragraph{\textbf{Electronic basis.}}
For the electrons the basis vectors are non-orthogonal Slater determinants built from atom-centered spin-orbitals, which are contractions of Gaussian type orbitals. Instead of a single Slater determinant per basis vector, one could also use valence bond structures (such as perfect pairing functions), which are eigenfunctions of the total spin operator and can be expanded into a linear combination of Slater determinants. 
Slater determinants are labeled with Greek lowercase letters. A Slater determinant is specified by fixing the occupancy of the spatial atomic orbitals and the spins (up or down) of the electrons in each orbital. 
\begin{equation}
\ket{\alpha(\Q)}
\end{equation}
Alternatively one can think that $\alpha$ encodes the way atomic orbitals are paired into Singlets, when one deals with valence bond structures instead of Slater determinants. In either case $\ket{\alpha(\Q)}$ is an analytical function of the nuclear coordinates $\Q$. 

$S_{\alpha,\beta}(\Q_i,\Q_j) = \bracket{\alpha(\Q_i)}{\beta(\Q_j)}$ is the overlap between two Slater determinants, in which the positions of the atomic orbitals differ. 
The Slater-Condon rules are not valid for Slater determinants built from overlapping orbitals. The rules that have to be used instead for 
${\bra{\alpha(\Q_i)} \hat{H}_{\text{elec}} \ket{\beta(\Q_j)}}$ - and in a slightly modified form also for \\
${\bra{\Q_i,\P_i} \otimes \bra{\alpha(\Q_i)} \hat{H}_{\text{elec-nuc}} \ket{\Q_j,\P_j} \otimes \ket{\beta(\Q_j)}  }$ - 
are well-known from valence bond theory\cite{matelems_verbeek, cooper_vb_theory}.
The matrix elements and their gradients make use of the 1st, 2nd and 3rd order cofactors of the overlap matrix $S_{\alpha,\beta}$ and of so-called compound matrices. The necessary mathematical background can be found in \cite{aitken}. Appendix \ref{sec:scaling} comments on the scaling of these operations.

\paragraph{\textbf{Moving basis set.}}
Now we come to the definition of the trajectories or moving basis vectors, which are tensor products of a nuclear coherent state and a linear combination of Slater determinants (or valence bond structures):
\begin{equation}
\ket{\Q_i,\P_i,\mathbf{a}_i} = \ket{\Q_i,\P_i} \otimes \sum_{A=1}^{N_{\text{st}}} a_i^A \ket{\chi_A(\Q_i)}
\end{equation}
where $\ket{\chi_A(\Q)}$ is an adiabatic eigenstate of the electronic hamiltonian for position $\Q$:
\begin{equation}
\ket{\chi_A(\Q)} = \sum_{\alpha=1}^{N_{\text{sl}}} U_{\alpha,A}(\Q) \ket{\alpha(\Q)}
\end{equation}
The $\chi_A(\Q)$ are real functions (in the absence of a magnetic field, which is not considered).

The columns of the matrix $\mathbf{U}$ contain the lowest few $N_{\text{st}}$ ($N_{\text{st}} < N_{\text{sl}}$) eigenvectors of the generalized eigenvalue problem
\begin{equation}
\sum_{\beta=1}^{N_{\text{sl}}} H^{\text{ord}}_{\alpha,\beta}(\Q,\P) U_{\beta,A}(\Q) = E_A(\Q,\P) \sum_{\beta=1}^{N_{\text{sl}}} S_{\alpha,\beta}(\Q) U_{\beta,A}(\Q) \quad \quad \quad A=1,\ldots,N_{\text{st}} \label{eqn:generalized_eigenproblem}
\end{equation}
with the adiabatic energies $E_A(\Q,\P)$. The indeces $\alpha,\beta$ labeling the matrix elements of the Hamiltonian
\begin{equation}
H^{\text{ord}}_{\alpha,\beta}(\Q,\P) = \bra{\Q,\P} \otimes \bra{\alpha(\Q)} \hat{H} \ket{\Q,\P} \otimes \ket{\beta(\Q)} 
\end{equation}
run over all $N_{\text{sl}}$ Slater determinants (or valence bond structures). The indeces $A$ and $B$ are used to enumerate the $N_{\text{st}}$ adiabatic electronic eigenstates.

Because of the averaging over a nuclear wavefunction, the adiabatic energies $E_A(\Q,\P)$ differ from the usual (?) adiabatic energies. In particular the singularity of the Coulomb potential for $\vert \mathbf{R}_a-\mathbf{R}_b \vert \to 0$ is removed.
Adiabatic electronic eigenstates belonging to the same nuclear coherent state $i$ are orthogonal:
\begin{equation}
\bracket{\chi_A(\Q_i)}{\chi_B(\Q_i)} = \delta_{AB}
\end{equation}
However, this is not the case for the overlap between eigenstates belonging to different nuclear coherent states:
\begin{equation}
\bracket{\chi_A(\Q_i)}{\chi_B(\Q_j)} = \sum_{\alpha,\beta=1}^{N_{\text{sl}}} U_{\alpha,A}^* \underbrace{\bracket{\alpha(\Q_i)}{\beta(\Q_j)}}_{S_{\alpha,\beta}(\Q_i,\Q_j)} U_{\beta,B}
\end{equation}

Gradients of the adiabatic energies, $\frac{\partial}{\partial \Q} E_A(\Q,\P)$ and non-adiabatic couplings, ${\bracket{\frac{\partial}{\partial \Q_i} \chi_A(\Q_i)}{\chi_B(\Q_j)}}$, can be reduced to gradients and gradient-couplings for non-orthogonal Slater determinants using the rules
for derivatives of eigenvectors and eigenvalues described in appendix \ref{sec:eigenderivatives}

\subsection{Time-Dependent Variational Principle}
Equations of motion for the nuclear phase space positions $\Q_i,\P_i$ and electronic amplitudes $\mathbf{a}_i$ can be motivated using the time-dependent variational principle\cite{karplus_wavepackets, fermion_molecular_dynamics}.
The Lagrangian for a normalized wavefunction $\ket{\Psi}$ reads:
\begin{equation}
L(\Psi,\dot{\Psi}) = \frac{\imath}{2} \left( \bracket{\Psi}{\dot{\Psi}} - \bracket{\dot{\Psi}}{\Psi} \right) - \bra{\Psi}\hat{H} \ket{\Psi}
\end{equation}
Assuming that $\ket{\Psi(t)} = \ket{\Q(t),\P(t),\mathbf{a}(t)}$ depends on time only through the parameters $\Q$,$\P$ and $\mathbf{a}$ the Lagrangian becomes:

\begin{equation}
\begin{split}
L & \left( \Q,\P,\mathbf{a},\frac{d\Q}{dt},\frac{d\P}{dt},\frac{d \mathbf{a}}{dt}\right) \\
 &= -\Q \cdot \frac{d \P}{dt} + \frac{\imath}{2} \sum_{A=1}^{N_{\text{st}}} \left( a^{A*} \frac{d a^A}{dt} - \frac{d a^{A*}}{dt} a^A \right)
                             - \imath \sum_{A,B=1}^{N_{\text{st}}} a^{A*} \bracket{\frac{\partial}{\partial \Q} \chi_A(\Q)}{\chi_B(\Q)} \cdot \frac{d \Q}{dt} a^B \\
                        & \quad \quad \quad     - \sum_{A=1}^{N_{\text{st}}} a^{A*} \left( E_A(\Q,\P) + V_{AB}^{\text{field}}(\Q,t)  \right) a^A 
\end{split}
\end{equation}
where $V_{AB}^{\text{field}}(\Q) = \bra{\chi_A(\Q)} \mathbf{E}(t) \cdot \mathbf{d} \ket{\chi_B(\Q)}$ describes the coupling between the external electric field and the (electronic) dipole operator.

The Euler-Lagrange equations
\begin{equation}
\frac{\partial L}{\partial \Q} = \frac{d}{dt} \frac{\partial L}{\partial \left(\frac{d \Q}{dt} \right)}, \quad 
\frac{\partial L}{\partial \P} = \frac{d}{dt} \frac{\partial L}{\partial \left(\frac{d \P}{dt} \right)}, \quad
\frac{\partial L}{\partial \mathbf{a}^*} = \frac{d}{dt} \frac{\partial L}{\partial \left(\frac{d \mathbf{a}^*}{dt} \right)}
\end{equation}
lead to
{\small
\begin{eqnarray}
  \frac{d \P}{dt} &=& -\sum_{A=1}^{N_{\text{st}}} a^{A*} \frac{\partial}{\partial \Q} \left( E_A(\Q,\P) + V_{AB}^{\text{field}}(\Q,t) \right) a^A + \imath \frac{d}{dt} \left( \sum_{A,B=1}^{N_{\text{st}}} a^{A*} \bracket{\frac{\partial}{\partial \Q} \chi_A(\Q)}{\chi_B(\Q)} a^B \right) \label{eqn:dPdt_Ehrenfest} \\
  \frac{d \Q_a}{dt} &=& \sum_{A,B=1}^{N_{\text{st}}} a^{A*} \frac{\partial}{\partial \P_a} E_A(\Q,\P) a^B = \frac{1}{M_a} \P_a \label{eqn:dQdt} \\
 \frac{d a^A}{dt} &=& - \imath \sum_{B} \left( E_A(\Q,\P) \delta_{AB} + V_{AB}^{\text{field}}(\Q,t) + \imath \bracket{\frac{\partial}{\partial \Q} \chi_A(\Q)}{\chi_B(\Q)} \cdot \frac{d \Q}{dt} \right) a^B \label{eqn:dadt}
\end{eqnarray}
}
In equation \ref{eqn:dPdt_Ehrenfest} the momentum change equals the state-averaged gradient plus a time-derivative of the non-adiabatic coupling vector.
The time-dependent variational principle leads to the equations for Ehrenfest dynamics, but we will be using surface hopping instead, which cannot be derived properly from a variational principle but allows for a more intuitive interpretation of results: The trajectory is propagated on a single adiabatic state $I$ (called the current state) and can switch stochastically between states, when the electronic amplitudes $\mathbf{a}$ decrease on the current state and increase on another one. Also, since the electric field oscillates very quickly, $V^{\text{field}}_{AB}$ is neglected and has no influence on the momentum change. The momentum change for surface hopping dynamics then reads:
\begin{equation}
\frac{d \P}{dt} = - \frac{\partial}{\partial \Q} E_I(\Q,\P)   \label{eqn:dPdt}
\end{equation}

In addition, the classical action is integrated along the path:
\begin{equation}
\frac{d S_i}{dt} = \sum_{a=1}^{N_{\text{at}}} \frac{1}{2} \left( \P_{a,i} \cdot \frac{d \Q_{a,i}}{dt} - \Q_{a,i} \cdot \frac{d \P_{a,i}}{dt} \right) \label{eqn:dSdt}
\end{equation}

\subsection{Coupling between trajectories}
The Schr\"{o}dinger equation
\begin{equation}
\imath \frac{\partial}{\partial t} \ket{\Psi} = \hat{H} \ket{\Psi} \label{eqn:schroedinger}
\end{equation}
is solved by
expanding the wavefunction into a moving basis set, with time-dependent coefficients $D_i(t)$:
\begin{equation}
\ket{\Psi} = \sum_{i=1}^{N_{\text{traj}}} \ket{\Q_i(t),\P_i(t),\mathbf{a}_i(t)} D_i(t) e^{\imath S_i(t)}
\end{equation}
The moving basis vectors (trajectories) will be labeled with the indeces $i$,$j$ and $k$. 
The basis is not orthonormal:
\begin{equation}
\Omega_{i,j} = \bracket{\mathbf{z}_i,\mathbf{a}_i}{\mathbf{z}_j,\mathbf{a}_j} = \bracket{\Q_i,\P_i,\mathbf{a}_i}{\Q_j,\P_j,\mathbf{a}_j} = \bracket{\Q_i,\P_i}{\Q_j,\P_j} \sum_{A,B=1}^{N_{\text{st}}} a_i^{A*} \bracket{\chi_A(\Q_i)}{\chi_B(\Q_j)} a_j^B
\end{equation}

Therefore the projection of the wavefunction onto a basis state
\begin{equation}
\bracket{\Q_i,\P_i,\mathbf{a}_i}{\Psi} = C_i(t) e^{\imath S_i} \label{eqn:projectionC}
\end{equation}
leads to another set of coefficients $C_i(t)$.

The expansion coefficients $C_i(t)$ and $D_i(t)$ are related by the matrix equation
\begin{equation}
\sum_k \Omega_{j,k} D_k e^{\imath S_k} = C_j e^{\imath S_j} \label{eqn:relation_D_and_C}
\end{equation}
By switching back and forth between the coefficients $C_i(t)$ and $D_i(t)$ during the integration the
calculation of the inverse overlap matrix can be avoided.

The time-dependence of $C_i(t)$ is found by taking the time-derivative of eqn. \ref{eqn:projectionC}. For simplicity we use the complex notation  $\ket{\mathbf{z}_i} = \ket{\Q_i,\P_i}$:
\begin{equation}
\begin{split}
\frac{d C_i(t)}{dt} e^{\imath S_i} =& \frac{d}{dt} \bracket{\mathbf{z}_i,\mathbf{a}_i}{\Psi} - \imath \frac{d S_i}{dt} \bracket{\mathbf{z}_i,\mathbf{a}_i}{\Psi} \\
 =& \left( \frac{d}{dt} \bra{\mathbf{z}_i,\mathbf{a}_i} \right) \ket{\Psi} + \bracket{\mathbf{z}_i,\mathbf{a}_i}{\frac{\partial}{\partial t} \Psi} - \imath \frac{d S_i}{dt} \bracket{\mathbf{z}_i,\mathbf{a}_i}{\Psi}
\end{split}
\end{equation}
We use the Schr\"{o}dinger equation \ref{eqn:schroedinger} to replace the time-derivative of the wavefunction and insert the discrete identity 
\begin{equation}
\identity = \sum_{j,k} \ket{\mathbf{z}_j,\mathbf{a}_j} \left( \Omega^{-1} \right)_{j,k} \bra{\mathbf{z}_k,\mathbf{a}_k}
\end{equation}
before $\ket{\Psi}$ to obtain:
\begin{equation}
\frac{d C_i(t)}{dt} e^{\imath S_i} = \sum_{j,k} \left\{
  \left( \frac{d}{dt} \bra{\mathbf{z}_i,\mathbf{a}_i} \right) \ket{\mathbf{z}_j,\mathbf{a}_j} 
- \imath \frac{d S_i}{dt} \bracket{\mathbf{z}_i,\mathbf{a}_i}{\mathbf{z}_j,\mathbf{a}_j} 
- \imath \bra{\mathbf{z}_i,\mathbf{a}_i} \hat{H} \ket{\mathbf{z}_j,\mathbf{a}_j} 
 \right\}
\times \left( \Omega^{-1} \right)_{j,k} \bracket{\mathbf{z}_k,\mathbf{a}_k}{\Psi}
\end{equation}
With the definition of the auxiliary expansion coefficients $D_i(t)$ in eqn. \ref{eqn:relation_D_and_C} we get:
\begin{equation}
\frac{d C_i(t)}{dt} e^{\imath S_i} = \sum_{j} \left\{ 
  \left( \frac{d}{dt} \bra{\mathbf{z}_i,\mathbf{a}_i} \right) \ket{\mathbf{z}_j,\mathbf{a}_j} 
- \imath \frac{d S_i}{dt} \bracket{\mathbf{z}_i,\mathbf{a}_i}{\mathbf{z}_j,\mathbf{a}_j} 
- \imath \bra{\mathbf{z}_i,\mathbf{a}_i} \hat{H} \ket{\mathbf{z}_j,\mathbf{a}_j} 
 \right\}
 D_j e^{\imath S_j} \label{eqn:C_kernel_D}
\end{equation}
The time-dependence of the basis vectors and the action gives
\begin{equation}
\begin{split}
\left(\frac{d}{dt} \bra{\mathbf{z}_i,\mathbf{a}_i} \right) & \ket{\mathbf{z}_j,\mathbf{a}_j} - \imath \frac{d S_i}{dt} \bracket{\mathbf{z}_i,\mathbf{a}_i}{\mathbf{z}_j,\mathbf{a}_j} \\
 =& \quad  \left\{ 
\left( \frac{d}{dt} \bra{\mathbf{z}_i} \right) \ket{\mathbf{z}_j} - \imath \frac{d S_i}{dt} \bracket{\mathbf{z}_i}{\mathbf{z}_j}
\right\} \sum_{A,B=1}^{N_{\text{st}}} a_i^{A*} \bracket{\chi_A(\Q_i)}{\chi_B(\Q_j)} a_j^B  \\
 & + \bracket{\mathbf{z}_i}{\mathbf{z}_j} \sum_{A,B=1}^{N_{\text{st}}} \frac{d a_i^{A*}}{dt} \bracket{\chi_A(\Q_i)}{\chi_B(\Q_j)} a_j^B \\
 & + \bracket{\mathbf{z}_i}{\mathbf{z}_j} \sum_{A,B=1}^{N_{\text{st}}} a_i^{A*} \bracket{\frac{\partial}{\partial \Q_i} \chi_A(\Q_i)}{\chi_B(\Q_j)} \cdot \frac{d \Q_i}{dt} a_j^B
\end{split} \label{eqn:dz_minus_dS}
\end{equation}

The term in curly braces in the first line of eqn. \ref{eqn:dz_minus_dS} is:
\begin{equation}
\left( \frac{d}{dt} \bra{\mathbf{z}_i} \right) \ket{\mathbf{z}_j} - \imath \frac{d S_i}{dt} \bracket{\mathbf{z}_i}{\mathbf{z}_j} = \bracket{\mathbf{z}_i}{\mathbf{z}_j} \left( \mathbf{z}_j - \mathbf{z}_i \right) \cdot \frac{d \mathbf{z}^*_i}{dt}
\end{equation}
In terms of the real vectors $\Q$ and $\P$ using the time-dependence of the action in eqn. \ref{eqn:dSdt}
we would have gotten the equivalent (but less compact) expression
\begin{equation}
\begin{split}
 \quad & \left( \frac{d}{dt} \bra{\Q_i,\P_i} \right) \ket{\Q_j,\P_j} - \imath \frac{d S_i}{dt} \bracket{\Q_i,\P_i}{\Q_j,\P_j} \\
=& \bracket{\Q_i,\P_i}{\Q_j,\P_j} \sum_{a=1}^{N_{\text{at}}} 
   \left( \sqrt{\alpha_a} \left(\Q_{a,j} - \Q_{a,i}\right) + \imath \sqrt{\frac{1}{4 \alpha_a}} \left(\P_{a,j} - \P_{a,i}\right) \right) \cdot
   \left( \sqrt{\alpha_a} \frac{d \Q_{a,i}}{dt} - \imath \sqrt{\frac{1}{4 \alpha_a}} \frac{d \P_{a,i}}{dt}  \right)
\end{split}
\end{equation}
where the time-derivatives of $\Q_i$ and $\P_i$ are calculated using the equations of motion \ref{eqn:dQdt} and \ref{eqn:dPdt}.

With the help of the time-dependence of the electronic amplitudes in \ref{eqn:dadt} one gets:

{\small
\begin{equation}
\begin{split}
\sum_{A,B=1}^{N_{\text{st}}} & \frac{d a_i^{A*}}{dt} \bracket{\chi_A(\Q_i)}{\chi_B(\Q_j)} a_j^B \\
& =  \imath \sum_{A,B=1}^{N_{\text{st}}} a_i^{A*} \left\{E_A(\Q_i,\P_i) \delta_{AB} + V_{BA}^{\text{field}}(\Q_i,t) + \imath \bracket{\frac{\partial}{\partial \Q_i} \chi_A(\Q_i)}{\chi_B(\Q_i)} \cdot \frac{d \Q_i}{dt} \right\} \sum_{C=1}^{N_{\text{st}}} \bracket{\chi_B(\Q_i)}{\chi_C(\Q_j)} a_j^C \label{eqn:dadt_chiAB_aB}
\end{split}
\end{equation}
}

Substituting eqn. \ref{eqn:dadt_chiAB_aB} into eqn. \ref{eqn:dz_minus_dS} and the result back into eqn. \ref{eqn:C_kernel_D} gives

\begin{equation}
\begin{split}
\frac{d C_i}{dt} e^{\imath S_i} = \sum_j \bracket{\mathbf{z}_i}{\mathbf{z}_j}  & \left\{ 
   \left( \mathbf{z}_j - \mathbf{z}_i \right) \cdot \frac{d \mathbf{z}_i^*}{dt} \sum_{A,B=1}^{N_{\text{st}}} a_i^{A*} \bracket{\chi_A(\Q_i)}{\chi_B(\Q_j)} a_j^B \right. \\
& \quad + \imath \sum_{A,B=1}^{N_{\text{st}}} a_i^{A*} \left( E_A(\Q_i,\P_i) \delta_{AB} + V^{\text{field}}_{BA}(\Q_i,t) \right) \sum_{C=1}^{N_\text{st}} \bracket{\chi_B(\Q_i)}{\chi_C(\Q_j)} a_j^C \\
& \quad - \imath \sum_{A,B=1}^{N_{\text{st}}} a_i^{A*} \left(
           H_{AB}^{\text{ord}}(\Q_i,\P_i;\Q_j,\P_j) + \delta_{ij} V_{AB}^{\text{field}}(\Q_i,t)
         \right) a_j^B \\
& \quad - \sum_{A,B=1}^{N_{\text{st}}} a_i^{A*} \bracket{\frac{\partial}{\partial \Q_i} \chi_A(\Q_i)}{\chi_B(\Q_i)} \cdot \frac{d \Q_i}{dt} \sum_{C=1}^{N_{\text{st}}} \bracket{\chi_B(\Q_i)}{\chi_C(\Q_j)} a_j^C \\
& \quad + \left. \sum_{A,B=1}^{N_{\text{st}}} a_i^{A*} \bracket{\frac{\partial}{\partial \Q_i} \chi_A(\Q_i)}{\chi_B(\Q_j)} \cdot \frac{d \Q_i}{dt} a_j^B \right\} D_j e^{\imath S_j} \label{eqn:kernel_long}
\end{split}
\end{equation}
where the adiabatic reordered hamiltonian is:
\begin{equation}
H_{AB}^{\text{ord}}(\Q_i,\P_i;\Q_j,\P_j) = \sum_{\alpha,\beta=1}^{N_{\text{sl}}} U_{\alpha,A}^{*} H_{\alpha,\beta}^{\text{ord}}(\Q_i,\P_i;\Q_j,\P_j) U_{\beta,B}
\end{equation}
For $i = j$, this matrix element reduces to the adiabatic eigenenergy $E_A$.

Above, we have made the approximation that the field does not couple different nuclear coherent states:
\begin{equation}
\bra{\chi_A(\Q_i)} \mathbf{E}(t) \cdot \mathbf{d} \ket{\chi_B(\Q_j)} = \delta_{ij} \bra{\chi_A(\Q_i)} \mathbf{E}(t) \cdot \mathbf{d} \ket{\chi_B(\Q_i)}
\end{equation}
This allows to take a larger time-step for the integration of the expansion coefficients $C_i(t)$ since the fast oscillating field only influences the electronic amplitudes $\mathbf{a}_i$. 

The terms in curly braces in eqn. \ref{eqn:kernel_long} define the kernel 
\begin{equation}
\delta^{2}\mathcal{H}(\Q_i,\P_i,\mathbf{a}_i; \Q_j,\P_j,\mathbf{a}_j) = -\imath \bracket{\mathbf{z}_i}{\mathbf{z}_j} \left\{ \ldots \right\}
\end{equation}
of the differential equation for the coefficients $C_i(t)$:
\begin{equation}
\frac{d C_i}{dt} e^{\imath S_i} = -\imath \sum_j \delta^2\mathcal{H} (\Q_i,\P_i,\mathbf{a}_i; \Q_j,\P_j,\mathbf{a}_j) D_j e^{\imath S_j} \label{eqn:dCdt}
\end{equation}

Note that the diagonal elements ($i = j$) of the kernel are zero, because in eqn. \ref{eqn:kernel_long} the second line cancels the third line and the fourth cancels the fifth line.

To make eqn. $\ref{eqn:kernel_long}$ more transparent the following abbreviations are introduced:
\begin{eqnarray}
\xi(i,j) &=& (\mathbf{z}_j - \mathbf{z}_i ) \cdot \frac{d \mathbf{z}^*_i}{dt} \label{eqn:xi} \\
O(i,j) &=&  \bracket{\mathbf{z}_i}{\mathbf{z_j}}
\end{eqnarray}
and the matrices
\begin{eqnarray}
S_{AB}(i,j) &=& \bracket{\chi_A(\Q_i)}{\chi_B(\Q_j)} \\
K_{AB}(i,j) &=& \bracket{\frac{\partial}{\partial \Q_i} \chi_A(\Q_i)}{\chi_B(\Q_j)} \cdot \frac{d \Q_i}{dt} \\
H_{AB}(i,j) &=& \sum_{\alpha,\beta=1}^{N_{\text{sl}}} U_{\alpha,A}^{*}(\Q_i) H_{\alpha,\beta}^{\text{ord}}(\Q_i,\P_i;\Q_j,\P_j) U_{\beta,B}(\Q_j) \label{eqn:H_AB}
\end{eqnarray}
with the electronic overlap, scalar non-adiabatic coupling and adiabatic hamiltonian between electronic eigenstates at different positions $\Q_i,\P_i$ and $\Q_j,\P_j$ in nuclear phase space. 
For $i \neq j$, the kernel coupling two different moving basis vectors can be written as:

\begin{equation}
\begin{split}
\delta^{2}\mathcal{H}(\Q_i,\P_i,\mathbf{a}_i; & \Q_j,\P_j,\mathbf{a}_j)  \\
= O(i,j)
 & \times \mathbf{a}_i^{*} \cdot \Big( 
     - \imath \xi(i,j) \matr{S}(i,j)  \\
 &  \quad \quad \quad + \left[ \matr{H}(i,i) + \imath \matr{K}(i,i) \right] \matr{S}(i,j)
                      - \left[ \matr{H}(i,j) + \imath \matr{K}(i,j) \right] \Big) \cdot \mathbf{a}_j
\end{split}
\end{equation}

\paragraph{\textbf{Trajectory averages.}}
Even if the full wavefunction were available, analysing it would be extremely complicated. Only 2D cuts through the probability density or reduced densities, where all but 2 coordinates are integrated out, can be visualized. 
Also the transformation of the wavefunction into internal coordinates (or some reaction coordinates), that are more meaningful than the cartesian ones, is not a straightforward computation.
Therefore we seek an interpretation of the simulation results solely in terms of the trajectories. The coupling between the trajectories furnishes each trajectory with a weight
\begin{equation}
n_i = C_i^* D_i
\end{equation}
Since all weights sum to 1, 
\begin{equation}
\bracket{\Psi}{\Psi} = \sum_{i=1}^{N_{\text{traj}}} n_i = 1,
\end{equation}
the expectation value of a quantity $f$ can be approximated as a weighted average over trajectories:
\begin{equation}
\langle f  \rangle = \sum_{i=1}^{N_{\text{traj}}} n_i f(\Q_i,\P_i,\mathbf{a}_i)
\end{equation}
As opposed to this, if the trajectories are propagated independently, the weights remain the same for all trajectories, i.e.
$n_i = \frac{1}{N_{\text{traj}}}$, and interference effects are neglected in the trajectory average.

An ensemble of trajectories can be visualized conveniently by superimposing the molecular geometries in a viewer. The weight can be shown by displaying trajectories with higher weight
more opaquely.


\section{Conclusion}
A method for simulating non-adiabatic dynamics based on valence bond theory has been presented, where matrix elements between Gaussians are calculated exactly, although the surfaces do not have to be calculated in advance. 

As a next step the method should be tested on a simple molecule,such as H$_3$.

Instead of atom-centered electronic orbitals, one could use coherent states of variable width for the electrons, as well. Such floating and breathing orbitals have been tested in \cite{corpuscular_electrons}.

\appendix
\section{Integrals of nuclear orbitals}
overlap
\begin{equation}
\bracket{\Q_{a,i},\P_{a,i}}{\Q_{a,j},\P_{a,j}} = \exp\left(
     -\frac{1}{2} \left[ 
                         \alpha_a \left(\Q_{a,i} - \Q_{a,j} \right)^2 
                        + \frac{1}{4 \alpha_a} \left( \P_{a,i} - \P_{a,j} \right)^2 
                        - \imath \left( \Q_{a,i} \cdot \P_{a,j} - \P_{a,i} \cdot \Q_{a,j}\right) 
                  \right] 
                \right)
\end{equation}
kinetic energy
{\small
\begin{equation}
\bra{\Q_{a,i},\P_{a,i}} - \frac{1}{2 M_a} \nabla_a^2 \ket{\Q_{a,j},\P_{a,j}} = 
      \frac{\alpha_a}{2 M_a} \left( 3 - \alpha_a \left[\Q_{a,i}-\Q_{a,j} - \frac{\imath}{2 \alpha_a} \left(\P_{a,i} + \P_{a,j}\right) \right]^2 \right) 
      \bracket{\Q_{a,i},\P_{a,i}}{\Q_{a,j},\P_{a,j}} 
\end{equation}
}
dipole matrix elements
{\small
\begin{equation}
\bra{\Q_{a,i},\P_{a,i}} \mathbf{R}_a \ket{\Q_{a,j},\P_{a,j}} = 
      \frac{1}{2} \left( \left(\Q_{a,i}-\Q_{a,j}\right) - \frac{\imath}{2 \alpha_a} \left(\P_{a,i} + \P_{a,j}\right) \right) 
      \bracket{\Q_{a,i},\P_{a,i}}{\Q_{a,j},\P_{a,j}} 
\end{equation}
}

repulsion between ions (nuclear equivalent of the two-electron integrals)
{\tiny
\begin{equation}
\begin{split}
\left( \Q_{a,i},\P_{a,i};\Q_{a,j},\P_{a,j} \vert \Q_{b,i},\P_{b,i}; \Q_{b,j},\P_{b,j} \right) &= \int d^3R_a \int d^3R_b \bracket{\Q_{a,i},\P_{a,i}}{\mathbf{R}_a} \bracket{\Q_{b,i},\P_{b,i}}{\mathbf{R}_b} \frac{Z_a Z_b}{\vert \mathbf{R}_a - \mathbf{R}_b \vert} \bracket{\mathbf{R}_a}{\Q_{a,j},\P_{a,j}} \bracket{\mathbf{R}_b}{\Q_{b,j},\P_{b,j}} \\
 = Z_a Z_b 2 \sqrt{\frac{2}{\pi} \frac{\alpha_a \alpha_b}{\alpha_a + \alpha_b}} & \bracket{\Q_{a,i},\P_{a,i}}{\Q_{a,j},\P_{a,j}} \bracket{\Q_{b,i},\P_{b,i}}{\Q_{b,j},\P_{b,j}} \\
                              \times  & F_0\left( 
                                \frac{\alpha_a \alpha_b}{\alpha_a + \alpha_b} \frac{1}{2} 
                                \left[ \left( \Q_{a,i} + \Q_{a,j} \right) - \frac{\imath}{2 \alpha_a} \left( \P_{a,i} - \P_{a,j} \right)
                                      -\left( \Q_{b,i} + \Q_{b,j} \right) + \frac{\imath}{2 \alpha_b} \left( \P_{b,i} - \P_{b,j} \right)
                                  \right]^2 \right)
\end{split}
\end{equation}
}
where
\begin{equation}
F_0(t) = \frac{1}{2} \sqrt{\frac{\pi}{t}} \erf\left(\sqrt{t}\right)
\end{equation}

mixed nuclear-electronic integrals (equivalents of the attractive nuclear integrals $(m|-\frac{Z_c}{\mathbf{r} - \mathbf{R}_c} |n)$ ):
The electronic atomic orbitals are composed of s-type Gaussian lobes only. p-orbitals are represented by two displaced s-orbitals with opposite signs. Therefore the primitive integrals, from which the electronic atomic orbitals are built by contraction, only involve s-type orbitals. $g_m(\mathbf{r} - \Q_{a,i})$ and $g_n(\mathbf{r} - \Q_{b,j})$ denote two primitive Gaussian s-orbitals centered at the position of the nuclear coherent states $\Q_{a,i}$ and $\Q_{b,j}$ with exponents $\alpha_m$ and $\alpha_n$:
\begin{equation}
g_m(r) = \left( \frac{2 \alpha_m}{\pi} \right)^{3/4} \exp\left(-\alpha_m \vert r \vert^2 \right)   \text{ normalized primitive Gaussian orbital }
\end{equation}

\begin{equation}
\begin{split}
\bra{\Q_{c,i},\P_{c,i}} & \otimes \bra{m(\Q_{a,i})} \frac{-Z_c}{\vert \mathbf{r} - \mathbf{R}_c \vert} \ket{\Q_{c,j},\P_{c,j}} \otimes \ket{n(\Q_{b,j})} \\
 =&  \int d^3R_a \int d^3r \bracket{\Q_{c,i},\P_{c,i}}{\mathbf{R}_a} g_m(\mathbf{r} - \Q_{a,i}) \frac{-Z_c}{\vert \mathbf{r} - \mathbf{R}_c \vert} g_n(\mathbf{r} - \Q_{b,j}) \bracket{\mathbf{R}_a}{\Q_{c,j},\P_{c,j}}  \\
 =& -Z_a 
 \left( \frac{2 \alpha_c}{\pi} \right)^{3/2} 
 \left( \frac{2 \alpha_m}{\pi} \right)^{3/4}
 \left( \frac{2 \alpha_n}{\pi} \right)^{3/4}
\frac{2 \pi^{5/2}}{(\alpha_m + \alpha_n)(2 \alpha_c) \sqrt{\alpha_m + \alpha_n + 2 \alpha_c}} \\
 & \times \exp\left(- \frac{\alpha_m \alpha_n}{\alpha_m + \alpha_n} \left(\Q_{a,i} - \Q_{b,j}\right)^2 \right)
       \bracket{\Q_{c,i},\P_{c,i}}{\Q_{c,j},\P_{c,j}} \\
 & \times F_0\left(
\frac{(\alpha_m + \alpha_n) (2 \alpha_c)}{\alpha_m + \alpha_n + 2 \alpha_c} 
             \left[
               \frac{1}{\alpha_m + \alpha_n} \left(\alpha_m \Q_{a,i} + \alpha_n \Q_{b,j} \right)
               -\frac{1}{2} \left( \Q_{c,i} + \Q_{c,j} - \frac{\imath}{2 \alpha_c} \left( \P_{c,i} - \P_{c,j} \right)\right)
             \right]^2
\right)
\end{split}
\end{equation}

\section{Derivatives of eigenvalues and eigenvectors}
\label{sec:eigenderivatives}
Generalized eigenvalue problem \ref{eqn:generalized_eigenproblem} (simplified notation)
\begin{equation}
\H(q) \U = \S(q) \U \E, \quad \U^{\dagger} \S \U = \identity \label{eqn:HUeqSUE}
\end{equation}
$q$ is a component of the vector $\Q$. $\E$ is a diagonal matrix with the eigenenergies.
$\H(q)$ and $\S(q)$ depend explicitly on the nuclear coordinate $q$ while the eigenvectors $\U$ and the eigenenergies $\E$ depend on them implicitly. All matrices $\in {\rm I\!R}^{N_{\text{sl}} \times N_{\text{sl}}}$. We need to find the gradients of $\E$ and $\U$ with respect to $q$. Since the eigenvectors are a basis for the vector space ${\rm I\!R}^{N_{\text{sl}}}$, 
\begin{equation}
\frac{\partial \U}{\partial q} = \U \C \label{eqn:dUeqUC}
\end{equation}
Taking the $q$-derivative of equation \ref{eqn:HUeqSUE}, substituting eqn. \ref{eqn:dUeqUC}, multiplying from the left with $\U^{\dagger}$ and using the fact that $\U^{\dagger} \H \U = \E$ gives:
\begin{equation}
\U^{\dagger} \frac{\partial \H}{\partial q} \U - \U^{\dagger} \frac{\partial \S}{\partial q} \U \E - \frac{\partial \E}{\partial q} = \C \E - \E \C
\end{equation}
Since $\E$ is a diagonal matrix, the diagonal elements of the right hand side of this equation are zero, so that one can solve for the gradient of the energy:
\begin{equation}
\frac{\partial E_A}{\partial \Q} = \sum_{\alpha,\beta=1}^{N_{\text{sl}}} U_{\alpha,A}^* \left(\frac{\partial H^{\text{ord}}_{\alpha,\beta}}{\partial \Q} - E_A \frac{\partial S_{\alpha,\beta}}{\partial \Q} \right) U_{\beta,A}
\end{equation}

Assuming that the lowest eigenvalues are not exactly degenerate, the off-diagonal elements of the coordinates of the gradient vectors are:
\begin{equation}
C_{AB} = \frac{1}{E_B-E_A} \sum_{\alpha,\beta=1}^{N_{\text{sl}}} U_{\alpha,A}^* \left( \frac{\partial H^{\text{ord}}_{\alpha,\beta}}{\partial \Q} - E_B \frac{\partial S_{\alpha,\beta}}{\partial \Q} \right) U_{\beta,B} \quad \quad \text{ for } A \neq B
\end{equation}
The diagonal elements follow from taking the derivative of $\U^{\dagger} \S \U = \identity$, which leads to
\begin{equation}
\C + \C^{\dagger} = - \U^{\dagger} \frac{\partial \S}{\partial q} \U 
\end{equation}
This equation can be solved for the diagonal elements of $\C$:
\begin{equation}
C_{AA} = - \frac{1}{2} \sum_{\alpha,\beta=1}^{N_{\text{sl}}} U_{\alpha,A}^* \frac{\partial S_{\alpha,\beta}}{\partial \Q} U_{\beta,A}
\end{equation}

In the case of degeneracy, the derivatives of the eigenvectors can be found using slightly more complicated formulae\cite{eigenderivatives}, which require the second derivatives of $\mathbf{H}^{\text{ord}}_{\alpha,\beta}$ and $S_{\alpha,\beta}$.

Knowing the eigenvectors and eigenvector derivatives for electronic hamiltonian at different coherent state positions one can compute the non-adiabatic couplings:
\begin{equation}
\bracket{\frac{\partial}{\partial \Q_i} \chi_A(\Q_i)}{\chi_B(\Q_j)} = \sum_{\alpha,\beta=1}^{N_{\text{sl}}} \left( \frac{\partial U_{\alpha,A}^{*}}{\partial \Q_i} \bracket{\alpha(\Q_i)}{\beta(\Q_j)}  + U_{\alpha,A}^{*} \bracket{\frac{\partial}{\partial \Q_i} \alpha(\Q_i)}{\beta(\Q_j)} \right) U_{\beta,B}(\Q_j)
\end{equation}

\section{Initial conditions for Gaussian wavepackets}

\section{Numerical integration scheme}
The numerical integration can be made more efficient by noting that the timescales for the motion of the heavy nuclei (sub-femtoseconds) and the lighter electrons (sub-attoseconds) are vastly different. 
During the integration the two timescales can be separated by writing

\begin{eqnarray}
\Omega(\Q_i,\P_i,\mathbf{a}_i; \Q_j,\P_j,\mathbf{a}_j) &=& \mathbf{a}_i^* \cdot \Omega_{\text{nuc}}(\Q_i,\P_i;\Q_j,\P_j) \cdot \mathbf{a}_j \label{eqn:Omega_elec} \\
\delta^{2}\mathcal{H}(\Q_i,\P_i,\mathbf{a}_i; \Q_j,\P_j,\mathbf{a}_j) &=& \mathbf{a}_i^* \cdot \delta^2\mathcal{H}_{\text{nuc}}(\Q_i,\P_i;\Q_j,\P_j) \cdot \mathbf{a}_j \label{eqn:d2H_elec}
\end{eqnarray}

where the matrices
\begin{eqnarray}
\Omega_{\text{nuc}}(\Q_i,\P_i;\Q_j,\P_j) &=& O(i,j) \matr{S}(i,j) \\
\delta^2\mathcal{H}_{\text{nuc}}(\Q_i,\P_i;\Q_j,\P_j) &=& O(i,j) \Big( 
     - \imath \xi(i,j) \matr{S}(i,j) \\
      && \quad \quad \quad + \left[ \matr{H}(i,i) + \imath \matr{K}(i,i) \right] \matr{S}(i,j)
                           - \left[ \matr{H}(i,j) + \imath \matr{K}(i,j) \right] \Big)
\end{eqnarray}
do not depend on the electronic amplitudes but only on the nuclear phase space position. 

$\xi(i,j)$,$O(i,j)$ and the matrices $\matr{S}(i,j)$, $\matr{K}(i,j)$ and $\matr{H}(i,j)$ (defined in eqns. \ref{eqn:xi}-\ref{eqn:H_AB}) only have to be calculated every nuclear timestep as the nuclear phase space positions $\Q_i$ and $\P_i$ change much slower than the electronic amplitudes $\mathbf{a}_i$ and the coefficients $C_i$. The same is true for the electronic Hamiltonian (the term in brackets in eqn. \ref{eqn:dadt}) that is used to integrate the electronic amplitudes. Therefore, the nuclear time step ($\Delta t_{\text{nuc}} = 0.1$ fs) is subdivided into $N_t = 1000$ electronic time steps and the mentioned quantities are interpolated linearly between the endpoints of the nuclear time interval. In this way, the expensive electronic structure calculation is needed only once every nuclear time step.
 
The integration of the electronic amplitudes and the coefficients is intertwined using a 4th order Runge-Kutta integrator. In each electronic time-step the overlap matrix and the coupling kernel are constructed according to eqns. \ref{eqn:Omega_elec} and \ref{eqn:d2H_elec} using the interpolated nuclear quantities and the electronic amplitudes and coefficients are advanced according to eqns. \ref{eqn:dadt} and \ref{eqn:dCdt}, respectively. A matrix equation has to be solved once in each of the 4 Runge-Kutta steps to switch between $C$ and the auxiliary coefficients $D$. 
The nuclear and electronic time steps are chosen small enough so that the total energy and electronic amplitude of each trajectory is conserved as well as the norm of the total wavefunction,  $\bracket{\Psi}{\Psi} = \sum_{i=1}^{N_{\text{traj}}} C_i^* D_i$.

\section{Scaling}
\label{sec:scaling}

The scaling of the algorithm depends on 
\begin{itemize}
\item the number of electrons $N_{\text{elec}}$ and nuclei $N_{\text{at}}$,
\item the size of the basis set for electronic atomic orbitals $N_{\text{ao}}$,
\item the number $N_{\text{sl}}$ of Slater determinants (or valence bond structures) built from these atomic orbitals,
\item and the number of electronic time steps $N_t$ per nuclear time step
\end{itemize}

For each pair of trajectories the primitive electronic and nuclear integrals have to be calculated (which will scale at most as $\mathcal{O}(N_{\text{ao}}^4)$). Then the matrix elements of the hamiltonian between all Slater determinants have to be evaluated and the gradients for the diagonal elements. This requires 1st and 2nd cofactors, which take $\mathcal{O}(N_{\text{elec}}^4)$ operations. The gradients require 3rd order cofactors, which take $\mathcal{O}(N_{\text{elec}}^6)$ operations. 

If the valence bond structures are expanded into Slater determinants, there will be on the order of $2^{N_{\text{elec}}}$ Slater determinants. The cost of this expansion is what makes valence bond theory so unattractive. There are, however, developments that try to avoid this costly step\cite{Tableau_functions,paired_permanents,algebrants}. 

Since in our approximation the nuclear wavefunction does not have to be antisymmetrized with respect to exchange of identical atoms, the time for nuclear integrals is negligible compared with the electronic ones.

In each electronic time step, the linear equation \ref{eqn:relation_D_and_C} has to be solved, which costs on the order of $\mathcal{O}(N_t \times N_{\text{traj}}^2)$ operations.


\end{document}